\documentclass{ws-ijmpe}

\begin{document}

\markboth{Authors' Names}{Instructions for Typing Manuscripts (The
Density Matrix Renormalization Group and the Nuclear Shell Model)}

\catchline{}{}{}{}{}

\title{The Density Matrix Renormalization Group and the Nuclear Shell Model\\
 }

\author{\footnotesize S. Pittel}

\address{Bartol Research Institute and Department of Physics and Astronomy, University of Delaware, \\
Newark, DE 19086, USA\\
pittel@bartol.udel.edu}

\author{B. Thakur}

\address{Bartol Research Institute and Department of Physics and Astronomy, University of Delaware, \\
Newark, DE 19086, USA\\
}

\author{N. Sandulescu}

\address{Institute of Physics and Nuclear Engineering,
                      76900 Bucharest, Romania\\
}

\maketitle

\begin{history}
\received{(received date)}
\revised{(revised date)}
\end{history}

\begin{abstract}
We summarize recent efforts to develop an
angular-momentum-conserving variant of the Density Matrix
Renormalization Group method into a practical truncation strategy
for large-scale shell model calculations of atomic nuclei.
Following a brief description of the key elements of the method,
we report the results of test calculations for $^{48}$Cr and
$^{56}$Ni. In both cases we consider nucleons limited to the 2p-1f
shell and interacting via the KB3 interaction. Both calculations
produce a high level of agreement with the exact shell-model
results. Furthermore, and most importantly, the fraction of the
complete space required to achieve this high level of agreement
goes down rapidly as the size of the full space grows.
\end{abstract}

\section{Introduction}

In the nuclear shell model, the low-energy structure of a nucleus
is traditionally described by assuming an inert doubly--magic core
and diagonalizing the effective nuclear hamiltonian in an active
space involving at most a few major shells. Despite the enormous
truncation inherent in this approach, the method can still only be
applied in very limited nuclear regimes. For sufficiently heavy
nuclei, for example, further truncation further is required to
reduce the number of shell-model configurations to a manageable
size.

An attractive truncation possibility is provided by the Density
Matrix Renormalization Group (DMRG), a method initially developed
for low-dimensional quantum lattices\cite{White}, and later
extended to finite Fermi systems. In the latter context, it has
been applied to the description of small metallic
grains\cite{Duke1}, to problems in quantum chemistry\cite{QC} and
to two-dimensional electrons in strong magnetic fields \cite{2D}.
The successes achieved in these various applications suggests that
it might also prove useful as a dynamical truncation strategy for
obtaining accurate approximate solutions to the nuclear shell
model.

The DMRG method involves a systematic inclusion of the degrees of
freedom of the many-body problem. When treating quantum lattices,
real-space sites are added iteratively. In finite Fermi systems,
these sites are replaced by single-particle levels. At each stage,
the system [referred to as a {\em block}] is enlarged to include
an additional site or level. This enlarged block is then coupled
to the rest of the system (the {\em medium}) giving rise to the
{\em superblock}. For a given eigenstate of the superblock (often
the ground state) or perhaps for a group of important eigenstates,
the reduced density matrix of the enlarged block in the presence
of the medium is constructed and diagonalized and those states
with the largest eigenvalues are retained.

This process of systematically growing the system and determining
the optimal structure within that enlarged block is carried out
iteratively, by sweeping back and forth through the sites, at each
stage using the results from the previous sweep to define the
medium. In this way, the process iteratively updates the
information on each block until convergence from one sweep to the
next is achieved. Finally, the calculations are carried out as a
function of the number of states retained in each block, until the
changes are acceptably small.

The traditional DMRG method, when applied in nuclei and elsewhere,
works in a simple product space, whereby the enlarged block is
obtained as a product of states in the block and the added site
and likewise the superblock is obtained as a product of states in
the enlarged block and the medium. In the context of nuclear
terminology, this is equivalent to working in the m-scheme.

A limitation of the traditional algorithm is that it does not
preserve symmetries throughout the iterative enlargement process.
Since the density matrix procedure involves a truncation at each
of the iterative stages, there is a potential to lose these
symmetries and the associated correlations. On this basis, we
proposed\cite{Review} the adoption of a strategy whereby angular
momentum is preserved throughout the iterative DMRG process.  This
method, called the JDMRG, was applied in nuclear physics for the
first time in the context of the Gamow Shell Model\cite{PloPlo}.
It was subsequently developed for application to the traditional
shell model by Pittel and Sandulescu\cite{PS}, where a first test
application to $^{48}Cr$ was reported.

We have now dramatically improved the JDMRG algorithm, to the
point where it can be applied to significantly heavier nuclei. In
this presentation, we report test results for the largest
calculations carried out to date using this method, for the
nucleus $^{56}Ni$.

An outline of the paper is as follows. In section II, we provide a
brief overview of the traditional DMRG method including a
discussion of the changes required to incorporate angular-momentum
conservation throughout. In Section III, we report improved
calculations for $^{48}Cr$ relative to those of ref. \cite{PS}.
Then in Section IV, we report our recently-obtained results for
$^{56}Ni$. Finally, in Section V we summarize the principal
conclusions of this work and outline several directions for future
study.

\section{An overview of the DMRG method}

\subsection{Truncation}

The DMRG method is based on an iterative inclusion of the degrees
of freedom of the problem, represented as a chain of sites on a
lattice. This is illustrated schematically in Figure 1 for a
system with $8$ ordered sites.

\begin{figure}[th]
\centerline{\psfig{file=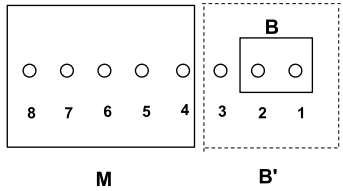,width=5cm}} \vspace*{8pt}
\caption{Schematic illustration of the DMRG growth procedure. A
  block $B$ consisting of sites $1$ and $2$ is enlarged to include site
  $3$, forming $B'$. The medium $M$ consists of all of the remaining
  sites, $4$ through $8$.}
\end{figure}

Assume that we have treated a group of these sites, referred to as
the block and denoted $B$, and that we have retained a total
number of states $m$ within that block.  We now wish to add to
this block the next site, which we assume contains $l$ states,
thereby producing an enlarged block $B'$. For the moment, we will
assume a product (or m-scheme) description, so that the enlarged
block has $m \times l$ states:

\begin{equation}
|i,j>_{B'}=|i>_B |j>_l ~~~, ~~~i=1,m ~~, ~~j=1,l
\end{equation}
As typical in Renormalization Group methods, we would like to
retain $m$ states for the enlarged block, exactly the same number
as before the enlargement. How do we choose them in an {\it
optimal} way?

In the DMRG method, we consider the enlarged block in the presence
of a medium $M$ that reflects all of the other sites of the
system, referring to the full system as the superblock (SB).
Assuming that the medium is likewise described by $m$ states, the
states of the superblock, $m \times l \times m$ in number, can be
expressed as
\begin{equation}
|i,j,k>_{SB} = |i,j>_{B'} |k>_M
\end{equation}

We then diagonalize the full hamiltonian of the system in the
superblock, for now isolating on its ground state,
\begin{equation}
|GS>_{SB}= \sum_{i,j,k} \Psi_{ijk} |i,j,k>_{SB} \label{GS}
\end{equation}
If we then construct the reduced density matrix of the enlarged
block in the ground state,
\begin{equation}
\rho_{ij,i'j'} = \sum_k \Psi^*_{ijk} \Psi_{i'j'k},
\end{equation}
diagonalize it and retain the $m$ eigenstates with the largest
eigenvalues we are {\it guaranteed} to have the $m$ most important
(or optimal) states of the enlarged block in the ground state
(\ref{GS}) of the superblock.

It is straightforward to target a group of states of the system
and not just the ground state. To do so,  we would construct a
mixed density matrix containing information on the block content
of all of them.

Once the optimal $m$ states are chosen, we renormalize all
required operators of the problem to the truncated space and store
this information. This would include all sub-operators of the
hamiltonian, viz:

\begin{displaymath}
a^{\dagger}_i, ~ a^{\dagger}_ia_j, ~ a^{\dagger}_ia^{\dagger}_j,
~a^{\dagger}_ia^{\dagger}_ja_k, ~a^{\dagger}_ia^{\dagger}_ja_ka_l,
~+~h.c.
\end{displaymath}
Having this information for the block and the additional level or
site enables us to calculate all such matrix elements for the
enlarged block as needed in the iterative growth procedure.

\subsection{Steps of the DMRG method}

With this as background, the DMRG procedure then involves the
following steps.

\subsubsection{Choice of an order for the sites}

Given a hamiltonian and the set of sites in which it is to act, we
need to define an order in which the sites are going to be
iteratively included. Depending on the number of sites, we might
wish to consider many (and perhaps all) orders to see which leads
to the energetically lowest solution. It is generally accepted
that the optimal order is one in which maximally entangled
sites/levels sit near one another\cite{Legeza1}.

\subsubsection{The warmup stage}

We begin the iterative process with what is traditionally called
the warmup stage. Here, we make an initial guess on the optimal
$m$ states for each possible block that will be constructed. This
choice will be important in determining how rapidly the iterative
procedure will converge. In our treatment, we do this by growing
blocks from each side of the chain gradually, using those orbits
that have already been treated on the other end as the medium.
This is illustrated in Figure 2 for the case of enlargement of the
right block from $2$ to $3$ sites, in the presence of a medium
involving the two-site left block already treated.

\begin{figure}[th]
\centerline{\psfig{file=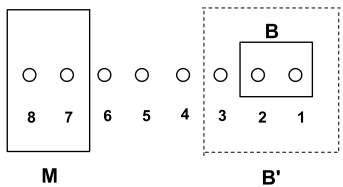,width=5cm}} \vspace*{8pt}
\caption{Schematic illustration of the warmup stage as used in the
calculations reported herein.}
\end{figure}

\subsubsection{The Sweep stage}

In this stage of the iterative process, we gradually sweep through
the sites of the chain, using for the medium the results either
from the warmup phase (during the first sweep) or from the
previous sweep stage. As suggested earlier, this sweeping process
is done over and over until convergence is achieved in the results
from one sweep to the next.

\subsubsection{As a function of $m$}

The warmup and sweep steps just described are typically done for a
given choice of $m$. The calculations are then carried out as a
function of $m$, until the changes with increasing $m$ are
acceptably small.

There is an alternative approach that has been proposed whereby a
given set of calculations are carried out not as a function of $m$
but rather as a function of the fraction of the complete density
matrix that is retained\cite{Legeza2}.

The calculations we will report are based on the former approach.

\subsection{The JDMRG approach}

As noted earlier, most DMRG approaches violate symmetries. In
nuclei, for example, they typically work in the m-scheme. When
imposing truncation in such a procedure, however, it is difficult
to ensure that the states retained contain all the components
required by the Clebsch Gordan series to build states of good
angular momentum. For this reason, we have chosen to develop an
angular-momentum-conserving variant of the DMRG method in which
angular momentum is preserved throughout the growth, truncation
and renormalization stages, referring to it as the JDMRG method.

The JDMRG approach follows the traditional DMRG approach outlined
above in most respects. The most significant change is that now we
must calculate and store throughout the iterative process the {\em
reduced matrix elements} of all sub-operators of the hamiltonian,
namely

\begin{eqnarray*}
&& a^{\dagger}_i, ~ [a^{\dagger}_i\tilde{a}_j]^K, ~
[a^{\dagger}_ia^{\dagger}_j]^K, ~\left( [a^{\dagger}_ia^{\dagger}_j]^K\tilde{a}_k\right)^L, \nonumber \\
&& ~\left([a^{\dagger}_ia^{\dagger}_j]^K
~[\tilde{a}_k\tilde{a}_l]^K \right)^0~ +h.c.
\end{eqnarray*}
This can be done using standard Racah algebra methods.

\subsection{A three-block JDMRG strategy}

In the calculations we will report, we adopt a three-block
strategy for the enlargement and truncation process. The basic
ideas are summarized in Figure 3.

We begin by choosing our order of sites so that neutron and proton
orbitals sit on opposite ends of the chain. We then gradually grow
blocks of each type of particle only, namely we grow neutron
blocks and proton blocks but no mixed blocks. Lastly, in the sweep
stage we go to and fro through the orbits of a given type of
particle only. As can be seen from the figure, the medium in this
approach involves two components. If, for example, we are
enlarging a proton block, as in the figure, the full medium ($M$)
involves all of the remaining proton levels and all of the neutron
levels.

This strategy, which was first proposed in ref. \cite{Review}, has
been found to be computationally more efficient than the two-block
approach that is customarily implemented when dealing with systems
involving two types of particles in which mixed blocks are
constructed.

\begin{figure}[th]
\centerline{\psfig{file=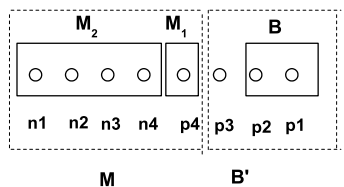,width=5cm}} \vspace*{8pt}
\caption{Schematic illustration of the three-block DMRG growth
procedure for a system with neutron and proton levels.}
\end{figure}

\section{Calculations}

We have carried out test calculations of this three-block JDMRG
approach on the nuclei $^{48}Cr$ and $^{56}Ni$. As in the usual
shell model approach, we assume that these nuclei can be described
in terms of valence neutrons and valence protons outside a
doubly-magic $^{40}Ca$ core. We use the shell-model hamiltonian
KB3\cite{KB3}, for which exact results are available for all
low-lying states in $^{48}Cr$\cite{Poves1} and for the
ground-state binding energy of $^{56}Ni$\cite{Poves2}.

We report the results separately for these two applications in the
following subsections.

\subsection{Results for $^{48}Cr$}

Here we discuss the results for $^{48}Cr$, where as noted earlier
preliminary results were presented in ref. \cite{PS}. The current
code, implemented on an Intel Xeon 5355 processor with 16 GB of
memory, runs significantly faster than the earlier calculations,
allowing us to test more features of the analysis. The size of the
full shell-model space for $^{48}Cr$ is 1,963,461 states. Of
these, 41,355 are $0^+$ states, 182,421 are $2^+$ states, etc.

These calculations are carried out by assuming an ordering of
single-particle levels as shown in Figure 4.

\begin{figure}[th]
\centerline{\psfig{file=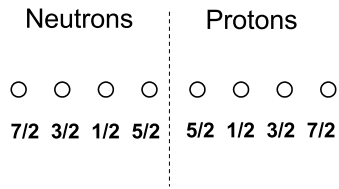,width=5cm}} \vspace*{8pt}
\caption{Order of single-particle levels assumed in all
calculations reported herein. }
\end{figure}

Our results for the ground state are presented in Table 1. The
exact calculation produces a ground state energy of $-32.953~MeV$.
The DMRG calculations converge smoothly to this result as $m$ is
increased, but require the inclusion of a substantial fraction of
the full space to obtain a high level of accuracy. For example,
with of order 25\% of the full $0^+$ space, we can achieve
accuracy to only a few $keV$. Even to achieve an accuracy of 50
$keV$, however, we require roughly 20\% of the full $0^+$ space.

It should be noted that the results reported in ref. \cite{PS}
were for a different order of single-particle levels, explaining
why the ground-state energies and maximum dimensions are different
for the same $m$ values.

\begin{table}[pt]
\tbl{Results for the energy of the ground state of $^{48}Cr$ in
$MeV$ from JDMRG calculations. {\em Max~Dim} refers to the maximum
dimension of the superblock hamiltonian and all results are given
in $MeV$. } {\begin{tabular}{@{}cccc@{}} \toprule $m$ &
$E_{GS}$ & Max~Dim \\
\colrule
\hphantom{000}40\hphantom{00} & \hphantom{0}-32.698 & \hphantom{00}1,985  \\
\hphantom{000}60\hphantom{00} & \hphantom{0}-32.763 & \hphantom{00}2,859  \\
\hphantom{000}80\hphantom{00} & \hphantom{0}-32.788 & \hphantom{00}3,765 \\
\hphantom{00}100\hphantom{00} & \hphantom{0}-32.789 & \hphantom{00}4,494  \\
\hphantom{00}120\hphantom{00} & \hphantom{0}-32.820 & \hphantom{00}6,367  \\
\hphantom{00}140\hphantom{00} & \hphantom{0}-32.843 & \hphantom{00}8,217 \\
\hphantom{00}160\hphantom{00} & \hphantom{0}-32.905 & \hphantom{00}9,948 \\
\hphantom{00}180\hphantom{00} & \hphantom{0}-32.945 & \hphantom{0}11,062 \\
\\
Exact   & \hphantom{0}-32.953 & \hphantom{0}41,355 \\
\botrule
\end{tabular}}
\end{table}

In Table 2, we present the corresponding results for the lowest
excited states, obtained for the blocks that derive in the ground
state optimization. We only show results for $m \geq 100$. Here
too convergence to the exact results is achieved, when a
sufficiently large fraction of the full shell-model space is
retained, with results of nearly comparable accuracy as for the
ground state. This is true despite the fact that the density
matrix truncation procedure implemented targeted the ground state
only.

\begin{table}[pt]
\tbl{Results for the energies of the lowest excited states in
$MeV$ from JDMRG calculations for $^{48}Cr$. {\em Dim} refers to
the dimension of the associated superblock hamiltonian and all
results are given in $MeV$. } {\begin{tabular}{@{}ccccc@{}}
\toprule $m$ &
$E_{2_1^+}$~~~(Dim) & $~E_{4_1^+}$~~~$(Dim)$ & $~E_{6_1^+}$~~~$(Dim)$& $~E_{0_2^+}$~~~$(Dim)$\\
\colrule

\hphantom{0}100& -31.977\hphantom{0} (21,003) & -30.897
\hphantom{0}(33,261)
&-29.157 \hphantom{0}(38,652) & -27.968 \hphantom{0}(4,494) \\
\hphantom{0}120 & -32.011 \hphantom{0}(28,677) & -30.935
\hphantom{0}(42,234)
&-29.200 \hphantom{0}(45,054)&-28.060 \hphantom{0}(6,367)\\
\hphantom{0}140 & -32.040 \hphantom{0}(36,706) & -30.979
\hphantom{0}(52,254)
&-29.264 \hphantom{0}(52,950)&-28.152 \hphantom{0}(8,217)\\
\hphantom{0}160 & -32.097 \hphantom{0}(44,454) & -31.042
\hphantom{0}(63,222)
& -29.341 \hphantom{0}(63,298)& -28.289 \hphantom{0}(9,948)\\
\hphantom{0}180 & -32.127 \hphantom{0}(50,030) & -31.087
\hphantom{0}(72,616)
& -29.433 \hphantom{0}(74,346)& -28.469 (11,062)\\
\\
Exact & -32.148 (182,421)& -31.130\hphantom{0}(246,979)
&-29.555 (226,259) &-28.564 (41,355)\\
\botrule
\end{tabular}}
\end{table}

\subsection{Results for $^{56}Ni$}

Next we turn to $^{56}Ni$, the largest calculation we have
performed to date. Now the size of the full shell-model space is
significantly larger. The full space in the m-scheme contains
1,087,455,228 states. In an angular momentum basis, the number of
$0^+$ states is 15,443,684. As noted earlier, we have only
considered $0^+$ states, since those are the only that were
calculated in a complete shell-model study with the KB3
interaction. Here too we assumed the order of single-particle
levels given in Figure 4.

The results for the ground-state energy as a function of $m$ are
shown in table 3. Here we are able to achieve roughly 60 $keV$
accuracy with barely 1\% of the full $0^+$ space.

\begin{table}[pt]
\tbl{Results for the energy of the ground state of $^{56}Ni$ in
$MeV$ from JDMRG calculations. {\em Max~Dim} refers to the maximum
dimension of the superblock hamiltonian and all results are given
in $MeV$. }{\begin{tabular}{@{}cccc@{}} \toprule $m$ &
$E_{GS}$ & Max~Dim \\
 \colrule
\hphantom{000}90\hphantom{00} & \hphantom{0}-78.360 & \hphantom{00}77,163 \\
\hphantom{00}120\hphantom{00} & \hphantom{0}-78.372 & \hphantom{0}102,690 \\
\hphantom{00}150\hphantom{00} & \hphantom{0}-78.388 & \hphantom{0}128,034 \\
\hphantom{00}180\hphantom{00} & \hphantom{0}-78.393 & \hphantom{0}162,019 \\
\\
Exact   & \hphantom{0}-78.46 & \hphantom{0}15,443,684 \\
\botrule
\end{tabular}}
\end{table}

This is perhaps the key result of the study. It suggests that the
fraction of the space required to achieve meaningful accuracy with
the JDMRG method goes down rapidly as the size of the space
increases. If this is confirmed with extension to even larger
problems it would bode very well for the future usefulness of the
JDMRG method as a practical truncation approach for large-scale
shell-model studies.

It should be noted that the DMRG was first applied to $^{56}Ni$ by
Papenbrock and Dean \cite{Papenbrock} in the traditional m-scheme.
Those calculations were unable to obtain the ground state energy
to better than $400$ $keV$.

As we were completing this study, we learned of a shell-model
treatment\cite{Horoi} of $^{56}Ni$ using the GXPF1A
interaction\cite{GXPF1A}. In that work, complete shell-model
results were reported for both the ground-state band and the first
deformed band. Clearly, it would be of interest for us to redo our
calculations for $^{56}Ni$ with this interaction, so that we could
test the convergence of our method in this nucleus for states
other than the ground state.

\section{Summary and Outlook}

In this talk, we have summarized the current status of our efforts
to build the Density Matrix Renormalization Group Method method
into a practical dynamical truncation strategy for large-scale
shell-model calculations of atomic nuclei. Following an overview
of the essential features of the method,  we discussed the changes
we had to implement for its use in application to nuclei. Most
importantly, we found it useful to develop an angular-momentum
conserving version of the method, the JDMRG. We then summarized
the principal results we have obtained to date with this method.
We reported test calculations for $^{48}$Cr and for $^{56}$Ni, in
both cases comparing with the results of exact diagonalization.
Both calculations were able to accurately reproduce the exact
shell-model results. In the case of $^{48}Cr$, however, this high
level of accuracy required us to retain a very large fraction of
the full space.  In contrast, we were able to achieve  accurate
results for $^{56}Ni$ with a much smaller fraction of the space.
The fact that the fraction of the space goes down with the size of
the problem is very promising for the future usefulness of the
method in even larger shell-model problems.

There are several issues that we intend to explore in the near
future. One concerns the need to determine through additional
calculations how rapidly the fraction of the space required for
convergence scales with the size of the problem. Currently we only
have two data points, $^{48}Cr$ and $^{56}Ni$. More are needed to
draw meaningful conclusions on this key point.

We also have some ideas as to how to break up large
single-particle orbitals while still preserving the angular
momentum invariance central to the JDMRG. We will test these ideas
in the context of the $f_{7/2}$ orbital in the current
applications and then look into their implementation for even
larger orbits.

As noted earlier, we also wish to repeat our test study of
$^{56}Ni$ for the GXPF1A interaction, where exact results also
exist for states other than the ground state.

Finally, we will wish to add the calculation of other observables,
including those connecting states in different systems.

Once these preliminaries have been completed, we expect to be in a
position for many interesting applications of the method to
problems of interest in nuclear structure physics.

\section*{Acknowledgements}

This work is based on a talk presented by one of the authors (SP)
at the International Workshop on Nuclear Structure Physics held in
Shanghai, China from 1-7 June 2008. It was supported by the US
National Science Foundation under grant \# PHY-0553127.  We
acknowldege with deep appreciation the important contributions of
Jorge Dukelsky to this project. We also wish to thank Alfredo
Poves for providing us with the KB3 matrix elements used in this
work.


\begin{thebibliography}{0}

\bibitem{White}  S.~R.~White, \emph{Phys. Rev. Lett.} \textbf{69}, 2863 (1992).

\bibitem{Duke1} J. Dukelsky and G. Sierra, Phys.
Rev. Lett. {\textbf 83}, 172 (1999);  J. Dukelsky and G. Sierra,
Phys. Rev. {\textbf B61}, 12302 (2000).

\bibitem{QC} S. R. White and R. L. Martin, J. Chem. Phys. {\textbf 110},
4127 (1999);  S. Daul, I. Ciofini, C. Daul, Steven R. White, Int.
J. Quantum Chem. {\textbf 79}, 331 (2000).

\bibitem{2D}
N. Shibata and D. Yoshioka, Phys. Rev. Lett. {\textbf 86}, 5755
(2001); D. Yoshioka and N. Shibata, Physica {\textbf E12}, 43
(2002); N. Shibata, J. Phys. A {\textbf 36}, R381 (2003).


\bibitem{Review} J.~Dukelsky, and S.~Pittel, Rep. Prog. Phys. \textbf{67}, 513 (2004).

\bibitem{PloPlo} N. Michel, W. Nazarewicz, M. Ploszajczak and J. Rotureau,
Rev. Mex. Fisica \textbf{50}, 74 (2004); J. Rotureau, N. Michel,
W. Nazarewicz, M. Ploszajczak and J. Dukelsky, Phys. Rev. Lett.
{\textbf 97}, 110603 (2006).

\bibitem{PS} S. Pittel and N. Sandulescu, Phys. Rev. C {\textbf 73}, 014301 (R)
(2006).

\bibitem{Legeza1} \"{O}. Legeza and J. S\'{o}lyom, Phys. Rev. B 68,
195116 (2003).

\bibitem{Legeza2} \"{O}. Legeza, J. R\"{o}der and B. A. Hess, Phys.
Rev. {\bf B67}, 125114 (2003).


\bibitem{KB3} A. Poves and A. P. Zuker, Phys. Rep. {\bf 70}, 235
(1981).

\bibitem{Poves1} E. Caurier, A. P. Zuker, A. Poves, G.
Mart\'inez-Pinedo, Phys. Rev. C {\textbf 50}, 225 (1994).

\bibitem{Poves2} E. Caurier, G. Mart\'inez-Pinedo, F. Nowacki, A.
Poves, J. Retamosa and A. P. Zuker, Phys. Rev. C {\textbf 59},
2033 (1999).

\bibitem{Papenbrock} T.~Papenbrock and D.~J.~Dean, J. Phys. G
{\textbf 31}, S1377 (2005).

\bibitem{Horoi} M. Horoi, B. A. Brown, T. Otsuka, M. Honma and T.
Mizusaki, Phys. Rev. C {\textbf 73}, 061305(R) (2006).

\bibitem{GXPF1A} M. Honma, T. Otsuka, B. A. Brown and T. Mizusaki,
Eur. Phys. J. A {\bf 25} Supp. 1, 499 (2005).



\end{thebibliography}
\end{document}